\documentstyle[aps,12pt]{revtex}

\begin{document}

\title{The Jacobi principal function in Quantum Mechanics}

\vskip2cm

\author{Rafael Ferraro{\footnote{e-mail:  ferraro@iafe.uba.ar}}}

\address{{\tighten
{\it Instituto de Astronom\'\i a y F\'\i sica del Espacio\\ Casilla de
Correo 67 - Sucursal 28\\1428 Buenos Aires, Argentina\\ and\\ Departamento
de F\'\i sica, Facultad de Ciencias Exactas y Naturales\\ Universidad de
Buenos Aires - Ciudad Universitaria, Pabell\'on I\\ 1428 Buenos Aires,
Argentina\\}}}

\maketitle

\vskip 1cm


\vskip1cm

\begin{abstract}
The canonical functional action in the path integral in phase space is 
discretized by linking each pair of consecutive vertebral points --${\bf 
q}_k$ and ${\bf p}_{k+1}$ or ${\bf p}_k$ and ${\bf q}_{k+1}$-- through the 
invariant complete solution of the Hamilton-Jacobi equation associated with 
the classical path defined by these extremes. When the measure is chosen to 
reflect the geometrical character of the propagator (it must behave as a 
density of weight 1/2 in both of its arguments), the resulting infinitesimal 
propagator is cast in the form of an expansion in a basis of short-time 
solutions of the wave equation, associated  with the eigenfunctions of the 
initial momenta canonically conjugated to a set of normal coordinates. The 
operator ordering induced by this prescription is a combination of a 
symmetrization rule coming from the phase, and a derivative term coming from 
the measure.
\end{abstract}  

\vskip  1cm

\narrowtext

\newpage

\section{Introduction}

By taking Dirac's ideas \cite{d} into account, R.P.Feynman explained how the
non-relativistic Quantum Mechanics can be formulated from principles that
make contact with the variational principles of the Lagrangian Mechanics\cite
{f}. Feynman showed that Quantum Mechanics can be based on the statement
that the {\it propagator}, i.e. the probability amplitude of finding the
system in the state ${\bf q}^{\prime \prime }$ at $t^{\prime \prime }$,
given that it was found in ${\bf q}^{\prime }$ at $t^{\prime }$, can be
obtained by means of the path integration: 
\begin{equation}
K({\bf q}^{\prime \prime }\ t^{\prime \prime }|{\bf q}^{\prime }\ t^{\prime
})\ =\ \int \ {\cal D}{\bf q}(t)\ \exp \left[ {\frac i\hbar }\ S[{\bf q}
(t)]\right] ,  \label{(1)}
\end{equation}
where $S[{\bf q}(t)]$ is the functional action of the system. Since the path
integral is a functional integration, one gives a meaning to Eq.(\ref{(1)})
by replacing each path by a {\it skeletonized} version where the path ${\bf q
}(t)$ is represented by a set of interpolating points $({\bf q}_k,t_k)$, $
k=0,1,...,N$, ${\bf q}_0={\bf q}^{\prime }$, ${\bf q}_N={\bf q}^{\prime
\prime }$, ${\bf q}_k={\bf q}(t_k)$. Then the functional action is replaced
by a function $S(\{{\bf q}_k,t_k\})$, and the functional integration reduces
to integrate the variables ${\bf q}_k$, $k=1,...,N-1$ \footnote{
The convergence is assured by endowing the time with an imaginary
part of proper sign.}. Finally the limit $\Delta t_k\equiv t_{k+1}-t_k
$ $\rightarrow 0$ (i.e., $N\rightarrow \infty $) is performed\footnote{
The rigorous mathematical meaning of this limit can be consulted in 
\onlinecite{albe} and references therein.}.

The function $S(\{{\bf q}_k,t_k\})$ is chosen to be\cite{f,dw}: 
\begin{equation}
S(\{{\bf q}_k,t_k\})=\sum_{k=0}^{N-1}S({\bf q}_{k+1}t_{k+1}|{\bf q}_kt_k)
\label{(2)}
\end{equation}
where $S({\bf q}_{k+1}t_{k+1}|{\bf q}_kt_k)$ is the Hamilton principal
function, i.e. the complete solution (in each argument) of the
Hamilton-Jacobi equation that is equal to the functional action evaluated on
the classical path joining its arguments. Thus the skeletonization (\ref{(2)})
 replaces each path ${\bf q}(t)$ by a succession of pieces defined by the
system itself, which join the interpolating points. The skeletonized action 
(\ref{(2)}) retains the essential classical property of the functional
action; namely it is stationary on the points interpolating the entire
classical path between $({\bf q}^{\prime },t^{\prime })$ and $({\bf q}
^{\prime \prime },t^{\prime \prime })$. In fact , $S(\{{\bf q}_k,t_k\})$ is
stationary when 
\begin{equation}
{\frac \partial {\partial {\bf q}_k}}S({\bf q}_{k+1}t_{k+1}|{\bf q}_kt_k)+{\ 
\frac \partial {\partial {\bf q}_k}}S({\bf q}_kt_k|{\bf q}_{k-1}t_{k-1})=0,\
\ \ \ \ \ \ \ \ \forall k  \label{(8)}
\end{equation}
meaning that the ${\bf q}_k$ are such that the final momentum of the
classical piece between $({\bf q}_{k-1},t_{k-1})$ and $({\bf q}_k,t_k)$,
matches the initial momentum of the classical piece between $({\bf q}_k,t_k)$
and $({\bf q}_{k+1},t_{k+1})$. This continuity guarantees that the points $
\{({\bf q}_k,t_k)\}$ are interpolating points of the entire classical path
between $({\bf q}^{\prime },t^{\prime })$ and $({\bf q}^{\prime \prime
},t^{\prime \prime })$.

\bigskip

Although a proper skeletonization for the path integral exists in the
configuration space, the measure in Eq.(\ref{(1)}) remains ambiguous. For
instance, the finite propagator for a quadratic Lagrangian is known to be 
\cite{fh,sch} 
\begin{equation}
K({\bf q}^{\prime \prime }t^{\prime \prime }|{\bf q}^{\prime }t^{\prime })\
=\ \left[ \det \left( {\frac i{2\pi \hbar }}{\frac{\partial ^2S({\bf 
q^{\prime \prime }}t^{\prime \prime }|{\bf q}^{\prime }t^{\prime })}{
\partial {\bf q}^{\prime \prime }\partial {\bf q}^{\prime }}}\right) \right]
^{1/2}\ \exp \left[ {\ \frac i\hbar }S({\bf q^{\prime \prime }}t^{\prime
\prime }|{\bf q}^{\prime }t^{\prime })\right] .  \label{(a0)}
\end{equation}
This expression is also valid for the infinitesimal propagator of any
classical system\cite{mor}. The prefactor in Eq.(\ref{(a0)}) is the Van
Vleck determinant\cite{vv}, which takes part in the measure, and
is nontrivial even for the short-time
pieces of the skeletonization. 
\bigskip

A different kind of example is the (finite) Newton-Wigner propagator for the
relativistic particle in flat space-time\cite{fe} 
\begin{equation}
K(q^{\prime \prime }\ t^{\prime \prime }|q^{\prime }\ t^{\prime })\ =\ -{\ 
\frac{(t^{\prime \prime }-t^{\prime })\ m^2\ c^3}{\pi \ \hbar \ S(q^{\prime
\prime }t^{\prime \prime }|q^{\prime }t^{\prime })}}\ K_1\left( {\frac i\hbar
}S(q^{\prime \prime }t^{\prime \prime }|q^{\prime }t^{\prime })\right)
\label{(a7)}
\end{equation}
(in 1+1 dimensions), where $S(q^{\prime \prime }t^{\prime \prime }|q^{\prime
}t^{\prime })=-mc\left( c^2(t^{\prime \prime }-t^{\prime })^2-(q^{\prime
\prime }-q^{\prime })^2\right) ^{1/2}$, and $K_1$ is a modified Bessel
function. In this case, the exponential of the Hamilton principal function
does not cleanly appear in the propagator, and neither does it in the short
time version (actually the propagator (\ref{(a7)}) gets the form (\ref{(a0)}
), not when $t^{\prime \prime }\rightarrow t^{\prime }$ but in the classical
limit when the Compton wavelength $\hbar /(mc)$ goes to zero). Results of
this sort could indicate a failure of (\ref{(1)}) to give the quantum
propagator for an arbitrary system\cite{and}. Anyway, it lays bare our
complete ignorance of the measure in the representation (\ref{(1)}).

\bigskip

It was thought that a path integration in phase space could remedy this
problem because there is a privileged measure in phase space: the Liouville
measure $d{\bf q}\ d{\bf p}\ /(2\pi\hbar)^n$ ($n$ is the dimension of the
configuration space), which is invariant under canonical transformations. In
this case, one should find an appropriate recipe for the skeletonization of
the canonical functional action 
\begin{equation}
S[{\bf q}(t),{\bf p}(t)]=\int_{t^{\prime}}^{t^{\prime\prime}} \left({\bf p}
(t)\cdot{\dot{{\bf q}}}(t) -H({\bf q},{\bf p})\right)\ dt,  \label{(3)}
\end{equation}
i.e., one should replace the functional in Eq.(\ref{(3)}) by a function $
S(\{ {\bf q}_k,{\bf p}_k,t_k\})$ of interpolating points for the path ${\bf 
q } (t), {\bf p}(t)$. The function $S(\{{\bf q}_k,{\bf p}_k,t_k\})$ should
be dictated by the system itself.

In Ref.\onlinecite{k} several recipes were essayed for a newtonian system
moving on a riemannian manifold. Since the data $({\bf q}_k,{\bf p}_k;{\bf 
q}_{k+1},{\bf p}_{k+1})$ overdeterminate the classical path between $t_k$ 
and $t_{k+1}$, then the basic idea was to use the classical piece in the
configuration space (just as in the previous case), together with the
parallel transport of ${\bf p}(t_k)$ $={\bf p}_k$ along that classical
piece. Of course, the parallel transport of ${\bf p}_k$ does not end in $
{\bf p}_{k+1}$ (unless the points interpolate the entire classical path
between $t^{\prime}$ and $t^{\prime\prime}$). So the skeletonized path
proves to be discontinuous in ${\bf p}$ (an unavoidable fact in phase
space). The different recipes for the skeletonization came from the
possibility of replacing the metric by a bitensor with the right coincidence
limit. After the momenta were integrated on, an infinitesimal propagator
similar to the one of Eq.(\ref{(a0)}) was obtained. However the different
skeletonizations reflected in a measure differing from the Van Vleck
determinant by corrections associated with the curvature of the manifold. As 
a consequence, the Hamiltonian operator in the Schr\"odinger equation had a 
term proportional to $\hbar^2 R$, where $R$ is the curvature scalar (see also 
Ref. \onlinecite{dw,p} for newtonian systems, and Ref.\onlinecite{fe} for
relativistic systems).

The skeletonization proposed in Ref.\onlinecite{k} successfully retains the
covariance of the system, but it does not treat coordinates and momenta on
an equal footing (a desirable feature in a canonical formalism).

\bigskip

In Ref.\onlinecite{fi} the use of complete solutions of the Hamilton-Jacobi
equation in the skeletonization has been suggested. A complete solution\cite
{ll,lanc} $\phi({\bf q},{\bf P},t)$ (the ${\bf P}^{\prime}s$ are $n$
integration constants) can be regarded as the generator of a canonical
transformation: ${\bf p} = \partial\phi/\partial{\bf q}$, ${\bf Q} =
\partial\phi/\partial{\bf P}$, where $({\bf Q},{\bf P})$ is a set of
classically conserved variables. Then $d\phi = {\bf p}\cdot{\bf dq} + {\bf Q}
\cdot d{\bf P} - H dt$, and ${\Delta\phi}_{{\bf P}}\equiv\phi({\bf q}
^{\prime\prime},{\bf P},t^{\prime\prime})$ $-$ $\phi({\bf q}^{\prime},{\bf P}
,t^{\prime})$ coincides with the canonical functional action evaluated along
a path such that ${\bf P}$ $= const$. Besides ${\Delta\phi}_{{\bf P}}$ is
stationary when ${\bf P}$ has the value corresponding to the
classical path joining $({\bf q}^{\prime},t^{\prime})$ and $({\bf q}
^{\prime\prime},t^{\prime\prime})$; in that case ${\Delta\phi}_{{\bf P}}$
turns out to be the Hamilton principal function\cite{lanc}. These properties
could make ${\Delta\phi}_{{\bf P}}$ a candidate to take part in the
skeletonization $S(\{{\bf q}_k,{\bf p}_k,t_k\})$.

But in order to obtain the propagator, some requirements concerned with the
behavior at short times and the character of the substitution ${\bf p}
\rightarrow{\bf P}$ --which must be well defined in all phase space--,
should be fulfilled by the complete solution to be chosen. In addition, the
canonical coordinates and momenta should enter the skeletonization on an
equal footing.

The rest of the paper is devoted to emphasize the role played in phase space
path integration by two related complete solutions of the Hamilton-Jacobi
equation, which will be called Jacobi principal functions. In Section II a
scheme of skeletonization putting canonical coordinates and momenta on an
equal footing suggests the initial condition that must be fulfilled by the
complete solutions to be used. In Section III the infinitesimal propagator
induced by the path integration is obtained once the measure is worked up
into a form that gives the same status to both arguments in the propagator.
Section IV explains how to treat a classical system with arbitrary
potential, in order to get the result (\ref{(a0)}) for the infinitesimal
propagator. Section V shows the ordering for the Hamiltonian operator that
is induced by the propagator of Section III. The conclusions are displayed
in Section VI.

\bigskip

\section{Skeletonization in phase space: the Jacobi principal function}

In order to introduce a skeletonization procedure treating coordinates and
momenta on an equal footing, one should define a recipe joining classical
pieces determined by mixed boundaries $({\bf q}_k,{\bf p}_{k+1})$ or $({\bf 
p }_k,{\bf q}_{k+1})$. Then a path ${\bf q}(t)$, ${\bf p}(t)$ should be
skeletonized by alternately giving the values of canonical coordinates and
momenta at each $t_k$, and replacing the canonical functional action by
something like 
\begin{equation}
S(\{{\bf q}_k,{\bf p}_k,t_k\})=\sum_{k=0}^{(N-2)/2}\left\{ J({\bf q}
_{2k+2}t_{2k+2}|{\bf p}_{2k+1}t_{2k+1})+J({\bf p}_{2k+1}t_{2k+1}|{\bf q}
_{2k}t_{2k})\right\} .  \label{(3')}
\end{equation}
The building blocks $J({\bf q\,}t^{\prime }|{\bf p\,}t)$ and $J({\bf p}
\,t^{\prime }|{\bf q\,}t)$ should be functions associated with the classical
system, making the skeletonized action stationary on the classical path in
phase space. A comparison with Eq. (\ref{(8)}) suggests that the stationary
condition 
\begin{equation}
{\frac \partial {\partial {\bf p}_{2k+1}}}J({\bf q}_{2k+2}t_{2k+2}|{\bf p}
_{2k+1}t_{2k+1})+{\frac \partial {\partial {\bf p}_{2k+1}}}J({\bf p}
_{2k+1}t_{2k+1}|{\bf q}_{2k}t_{2k})=0,\ \ \ \ \ \ \ \ \ \forall k
\label{(81)}
\end{equation}
should mean that the final canonical coordinates of the classical piece
between $t_{2k}$ and $t_{2k+1}$ coincides with the initial canonical
coordinates of the classical piece between $t_{2k+1}$ and $t_{2k+2}$. Once
the stationary value for the momenta is replaced in Eq.(\ref{(3')}), the
skeletonization should go to the one of Eq.(\ref{(2)}), so guaranteeing the
continuity of both ${\bf q}$ and ${\bf p}$ at $t_{2k+1}$.

If the system exhibits invariance under a general coordinate change, then the
skeletonization and the measure must preserve that invariance, in order that
the quantization is independent of the chosen coordinates.
Therefore both $J$'s in Eq.(\ref{(3')}) should be invariant.

\medskip\ 

We are going to define $\partial J/\partial {\bf p}$ to be the coordinate
canonically conjugated to ${\bf p}$; so we should look for coordinates
transforming contravariantly to ${\bf p}$. We will assume that the
configuration space is a riemannian manifold ${\cal M}$; thus, normal
coordinates ---which transform like the components of a vector at the origin
of coordinates---could be introduced\footnote{
Actually, only a connection is needed to define normal coordinates.}. Let us
choose a point $O\in {\cal M}$ as the origin of normal coordinates. Let 
$\{{\bf e}_a\}$ be a basis for the tangent space $T_O$ at $O$. To assign normal
coordinates to a point $P$, consider the geodesic joining $O$ and $P$ 
\footnote{We assume a global topology such that any pair of points is joined 
by a unique geodesic.} and define ${\bf \sigma} = s\,{\bf u}$, where $s=\int 
\sqrt{ g_{ij}dq^idq^j}$ is the (invariant) length of the geodesic between $O$
and $P $, and ${\bf u\in }T_O$ is the unitary vector tangent to the geodesic
at $O $ . The components $\sigma ^a(q^j)$ of the vector ${\bf \sigma }$ $\in 
$ $T_O $ are normal coordinates for $P$\cite{schutz}. By differentiating the
invariant Hamilton principal function with respect to the normal coordinates
of the (initial) final boundary, one gets the (initial) final momenta $
(-)\,p_a=\partial S/\partial \sigma ^a$; thus $p_a$ are the components
of a form ${\bf p}=p_a{\bf e}^a$ $\in $ $T_O^{*}$ . The set of canonically
conjugated variables $\{(\sigma ^a,$ $p_a)\}$ is invariant under changes $
q^j\rightarrow q^{j^{\prime }}$; $(\sigma ^a,$ $p_a)$ only change under
changes of the basis $\{{\bf e}_a\}$ in $T_O$ (and its dual basis $\{{\bf e}
_{\ }^a\}$ in $T_O^{*}$).

\bigskip

Now we will introduce two invariant Legendre transforms of the Hamilton
principal function $S({\bf q}^{\prime \prime }\ t^{\prime \prime }|{\bf q}
^{\prime }\ t^{\prime })$: 
\begin{equation}
J({\bf q}^{\prime \prime }\ t^{\prime \prime }|{\bf p}^{\prime }\ t^{\prime
})\ \equiv \ \left( S-{\frac{\partial S}{\partial \sigma ^{\prime }{}^a\ }}
\;\sigma ^{\prime }{}^a\right) _{p_a^{\prime }=-\partial S/\partial \sigma
^{\prime }{}^a}\;,  \label{(14)}
\end{equation}
and

\begin{equation}
J({\bf p}^{\prime \prime }\ t^{\prime \prime }|{\bf q}^{\prime }\ t^{\prime
})\ \equiv \ \ \left( S-{\frac{\partial S}{\partial \sigma ^{\prime \prime
}{}^a\ }}\;\sigma ^{\prime \prime }{}^a\right) _{p_a^{\prime \prime
}=\partial S/\partial \sigma ^{\prime \prime }{}^a}\;,  \label{(14')}
\end{equation}
that will be called {\it Jacobi principal functions}. They can be regarded
as the evaluation on the classical trajectory of a functional action that
has been added with surface terms to make it stationary under variations
with mixed boundaries left fixed.

\bigskip

In spite of their definition in terms of the Legendre transform
interchanging $\sigma ^a\ $and $p_a\ $, the Jacobi principal functions can
be written, if preferred, as functions of a different set of canonical
coordinates connected with the normal ones by means of a canonical point
transformation

\[
q^j=q^j\left( \sigma ^a\right) ,\;\;\;\;\;\;\;p_j=\frac{\partial \sigma ^a}{
\partial \,q^j}\;p_a\;. 
\]
The $p_j$ transform like the components of a form ${\bf p}\in $ $T_P^{*}$
on the coordinate basis .

\bigskip

From the properties of the Legendre transform, one easily gets that $J({\bf 
q }^{\prime \prime }\ t^{\prime \prime }|{\bf p}^{\prime }\ t^{\prime })$
and $-J({\bf p}^{\prime \prime }\ t^{\prime \prime }|{\bf q}^{\prime }\
t^{\prime })$ generate contact transformations:$\ $

\begin{equation}
p_j^{\prime \prime }\ =\ {\frac{\partial J({\bf q}^{\prime \prime }t^{\prime
\prime }|{\bf p}^{\prime }t^{\prime })}{\partial \,q^{\prime \prime \;j}\ }}
\ \ ,\ \ \ \ \ \ \ \ \ \ \ \ \ \ \ \ \ \ \ \ \ \ \sigma {^{\prime }{}^a}\ =\ 
{\ \frac{\partial J({\bf q}^{\prime \prime }t^{\prime \prime }|{\bf p}
^{\prime }t^{\prime })}{\partial \,p_a^{\prime }}}.  \label{(5')}
\end{equation}

and

\begin{equation}
\sigma ^{\prime \prime }{}^a\ =\ -{\frac{\partial J({\bf p}^{\prime \prime
}t^{\prime \prime }|{\bf q}^{\prime }t^{\prime })}{\partial \,p_a^{\prime
\prime }}},\ \ \ \ \ \ \ \ \ \ \ \ \ \ \ \ \ \ \ \ p_j^{\prime }\ =\ -{\frac{
\partial J({\bf p}^{\prime \prime }t^{\prime \prime }|{\bf q}^{\prime
}t^{\prime })}{\partial \,q^{\prime \;j}}}.  \label{(52')}
\end{equation}

\bigskip

When $t^{\prime \prime }=t^{\prime }$ they generate the identity:

\begin{equation}
J({\bf q\,}t\,|{\bf p\,}t)=p_a\ \sigma {}^a(q^j),  \label{(181)}
\end{equation}

\begin{equation}
J({\bf p\,}t\,|{\bf q\,}t)=-\ p_a\ \sigma {}^a(q^j)\ .  \label{(182)}
\end{equation}
\bigskip

$J({\bf q}^{\prime \prime }\ t^{\prime \prime }|{\bf p}^{\prime }\ t^{\prime
})$ and $-J({\bf p}^{\prime \prime }\ t^{\prime \prime }|{\bf q}^{\prime }\
t^{\prime })$ are complete solutions of the Hamilton-Jacobi equation in both
arguments (take $\partial /\partial t^{\prime \prime }$ and $\partial
/\partial t^{\prime }$ in Eqs. (\ref{(14)}) and (\ref{(14')}), and use Eqs. ( 
\ref{(5')}) and (\ref{(52')})):

\begin{equation}
\frac{\partial J({\bf q}\ t^{\prime \prime }|{\bf p}\ \ t^{\prime })}{
\partial \ t^{\prime \prime }}=-H(q^j{\bf \ },\frac{\partial J}{\partial
\,q^j{\bf \ }},t^{\prime \prime }),\ \ \ \ \ \ \ \ \ \ \ \ \frac{\partial J( 
{\bf q}\ t^{\prime \prime }|{\bf p}\ t^{\prime })}{\partial \ t^{\prime }}
=H(\sigma {}^a\!\!\!\,=\,\!\!\!\frac{\partial J}{\partial \ p_a{\bf \ }}
,\,p_j\,,t^{\prime })  \label{j1}
\end{equation}

\begin{equation}
-\frac{\partial J({\bf p}\ t^{\prime \prime }|{\bf q}\ t^{\prime })}{
\partial \ t^{\prime \prime }}=H(\sigma {}^a\!\!\!\,=\,\!\!\!-\frac{\partial
J}{\partial \ p_a{\bf \ }}\ ,\,p_j\,,t^{\prime \prime }),\ \ \ \ \ \ \ \ \ \
\ \frac{\partial J({\bf p}\ t^{\prime \prime }|{\bf q}\ t^{\prime })}{
\partial \ t^{\prime }}=H(q^j{\bf \ },-\frac{\partial J}{\partial {\bf \,}
q^j },t^{\prime })  \label{j2}
\end{equation}

\bigskip

By changing $t^{\prime \prime }\longleftrightarrow t^{\prime }$ in Eqs. (\ref
{j1}) and (\ref{j2}) one realizes that

\begin{equation}
J({\bf p}\;t^{\prime \prime }|{\bf q}\;t^{\prime })\ =\ -J({\bf q}
\;t^{\prime }|{\bf p}\;t^{\prime \prime }).  \label{(7')}
\end{equation}

\bigskip

If the system is conservative, then the Jacobi principal functions depend on 
$t^{\prime \prime }$ and $t^{\prime }$ only through the difference $
t^{\prime \prime }-t^{\prime }$. Therefore $J({\bf q}t^{\prime }|{\bf p}
t^{\prime \prime })\,=J({\bf q},{\bf p,\tau \equiv \Delta }t)$, and the
pieces of the skeletonized action (\ref{(3')}) have the form

\begin{eqnarray}
J({\bf q}_{2k+2}t_{2k+2}|{\bf p}_{2k+1}t_{2k+1}) &&+J({\bf p}
_{2k+1}t_{2k+1}| {\bf q}_{2k}t_{2k})  \nonumber \\
&=&J({\bf q}_{2k+2},{\bf p}_{2k+1},\ \tau _{2k+1})-J({\bf q}_{2k},{\bf p}
_{2k+1},-\ \tau _{2k})  \label{(91'')}
\end{eqnarray}
Thus the skeletonization (\ref{(3')}) gets the form ${\Delta \phi }_{{\bf P}
} $ proposed in Ref.\onlinecite{fi}, although a specific complete solution
of the Hamilton-Jacobi equation is being used here.

\medskip\ 

Even for a non conservative system, the short time limit of the
skeletonization (\ref{(3')}) is

\begin{eqnarray}
J({\bf q}_{2k+2}t_{2k+2} &&|{\bf p}_{2k+1}t_{2k+1})+J({\bf p}
_{2k+1}t_{2k+1}| {\bf q}_{2k}t_{2k})  \nonumber \\
&\simeq &\left[ {p_a}_{2k+1}{\ \sigma ^a}_{2k+2}-H({\ \sigma ^a}_{2k+2},{p_j}
_{2k+1},t_{2k+1})\,\Delta t_{2k+1}\right]  \nonumber \\
&&-\left[ {p_a}_{2k+1}{\ \sigma ^a}_{2k}+H({\ \sigma ^a}_{2k},{p_j}
_{2k+1},t_{2k+1})\,\Delta t_{2k}\right] \   \label{(91')}
\end{eqnarray}

\medskip\ 

On any smooth path it is valid that $\sigma _{2k+2}^a\rightarrow \ \sigma
_{2k}^a$ when $t_{2k+2}\rightarrow \ t_{2k}$ . Thus the skeletonized action
goes to the canonical functional action:

\begin{eqnarray}
J({\bf q}_{2k+2}t_{2k+2}\ &&|{\bf p}_{2k+1}t_{2k+1})+J({\bf p}
_{2k+1}t_{2k+1}| {\bf q}_{2k}t_{2k})  \nonumber \\
&&\longrightarrow \ {\ p_a}_{2k+1}\ (\sigma _{2k+2}^a-\ \sigma _{2k}^a)\ -H({
\ \ \sigma ^a}_{2k},{p_j}_{2k+1},t_{2k+1})\,(t_{2k+2}-\ t_{2k}).
\label{(91)}
\end{eqnarray}

\bigskip\ 

The skeletonization scheme proposed in this Section is based on a pair of
complete solutions of the Hamilton-Jacobi equation (in both arguments) that
treat canonical coordinates $\sigma ^a$ and momenta $p_a$ on an equal
footing; this fact is evident in the initial conditions (\ref{(181)}) and
(\ref {(182)}). The Jacobi principal functions do not depend nor on the chart $
\{q^j\}$ neither on the basis of the tangent space at $O$. They do depend on
the way the configuration space has been cut from the phase space (the
quantum propagation is not invariant under arbitrary canonical
transformations).

\section{The propagator}

In order to give sense to the functional integration in phase space

\begin{equation}
K({\bf q}^{\prime \prime }\ t^{\prime \prime }|{\bf q}^{\prime }\ t^{\prime
})\ =\ \int {\cal D}{\bf p}(t)\ {\cal D}{\bf q}(t)\ \exp \left[ {\frac i\hbar
}\ S[{\bf q}(t),{\bf p}(t)]\right] ,  \label{(4)}
\end{equation}
our attention must now turn on the measure. Since the functional action is
going to be replaced by an invariant skeletonized version, the ``magical ''
measure ${\cal D}{\bf p}(t)\ {\cal D}{\bf q}(t)\ $ should be consequently
replaced by a measure in the space of the variables $\{{\bf q}_{2k},{\bf 
p}_{2k+1}\}$. This measure must be able to retain the geometrical behavior
of the propagator, which is apparent in the manner of propagating the 
wavefunction:

\begin{equation}
\Psi ({\bf q}^{\prime \prime },t^{\prime \prime })=\int d{\bf q}^{\prime }\
K({\bf q}^{\prime \prime }t^{\prime \prime }|{\bf q}^{\prime }t^{\prime })\
\Psi ({\bf q}^{\prime },t^{\prime }).  \label{(92)}
\end{equation}
If the wave function $\Psi $ is regarded as scalar, then the propagator
should be invariant in its final argument but a density in its initial
argument. However an scalar wavefunction would compel us to use an
invariant measure $\mu ({\bf q})\ d{\bf q}$ in the inner product of the
Hilbert space (the density $\mu $ would be ultimately dictated by the result
of the path integration \cite{fe}). So it may be more convenient to regard
the wave function as a density of weight $1/2$. In this case the inner
product of the Hilbert space is

\begin{equation}
(\Psi ,\Phi )=\int d{\bf q}\ \Psi ^{*}\ \Phi ,  \label{(10)}
\end{equation}
whatever the generalized coordinates describing the system are. Thus the
propagator in Eq.(\ref{(92)}) must be a density of weight $1/2$ in both
arguments.

\smallskip

The issue of the measure can be studied at the level of an infinitesimal
propagator. In fact, due to the composition law

\begin{eqnarray}
K({\bf q}^{\prime \prime }\ t^{\prime \prime }|{\bf q}^{\prime }\ t^{\prime
})\ =\int &&K({\bf q}^{\prime \prime }t^{\prime \prime }|{\bf q}
_{N-1}t_{N-1})\ d{\bf q}_{N-1}\ K({\bf q}_{N-1}t_{N-1}|{\bf q}
_{N-2}t_{N-2}).......  \nonumber \\
&&\ \ .......\ d{\bf q}_2\ K({\bf q}_2t_2|{\bf q}_1t_1)\ d{\bf q}_1\ K({\bf 
q }_1t_1|{\bf q}^{\prime }t^{\prime })  \label{(27')}
\end{eqnarray}
---which holds whenever the added paths go forward in time---, the finite
propagator can be retrieved by composing infinitesimal propagators. If $
t^{\prime \prime }-t^{\prime }=\varepsilon $ is infinitesimal, then one
should only integrate ${\bf p}$ at some intermediate time $t$. However the
measure $d{\bf p}^{(t)}$ does not allow for a propagator behaving like a
density in its arguments ${\bf q}^{\prime }$ and ${\bf q}^{\prime \prime }$
. The use of $d^np_j^{\prime }\,=\left| \det \frac{\partial \,p_j^{\prime }}{
\partial \,p_a^{(t)}}\right| \,d^np_a^{(t)}\,$ instead of $d{\bf p}^{(t)}$
is suitable when the wave function is regarded as a scalar, because the
propagator will result a density in ${\bf q}^{\prime }$ . However, if the
propagator has to be a density of weight $1/2$ in both arguments, then the
Jacobian in the previous measure must be splitted in two factors that will
give an equal weight to ${\bf p}^{\prime }$ and ${\bf p}^{\prime \prime }$ :

\begin{eqnarray}
\left| \det \frac{\partial \,p_j^{\prime \prime }}{\partial \,p_a^{(t)}}
\right| ^{\ 1/2}\;&&d^np_a^{(t)}\;\left| \det \frac{\partial \,p_j^{\prime } 
}{\partial \,p_a^{(t)}}\right| ^{\ 1/2}  \nonumber \\
&&  \nonumber \\
&&=\left| \ \!{}\det \left( \frac{\partial ^2J({\bf q}^{\prime \prime
}\,t^{\prime \prime }\,|\,{\bf p\,}t)}{\partial q^{\prime \prime
\,j}\;\partial \,p_a}\right) \right| ^{\ 1/2}\;d^np_a^{(t)}\;\left| \ \!\det
\left( -{\frac{\partial ^2J({\bf p\,}t\,|\,{\bf q}^{\prime }\,t^{\prime })}{
\partial \,q^{\prime \,\,j}\;\partial \,p_a\;}}\right) \right| ^{\ 1/2}\ \ \
\label{(76)} 
\end{eqnarray}

\bigskip

Concretely, the infinitesimal propagator has the form:

\begin{eqnarray}
K({\bf q}^{\prime \prime }\ t^{\prime \prime }=t^{\prime }+\epsilon \;|\; 
{\bf q}^{\prime }\ t^{\prime })=\int {\frac{d^np_a}{(2\pi \hbar )^n}}\
&&\left| \ \!{}\det \frac{\partial ^2J({\bf q}^{\prime \prime }\,t^{\prime
\prime }\,|\,{\bf p\,}t)}{\partial \,q^{\prime \prime \,\,j}\;\partial
\,p_a\;}\ \right| ^{\ 1/2\ }\;\left| \ \!\det -{\frac{\partial ^2J({\bf p\,}
t\,|\,{\bf q}^{\prime }\,t^{\prime })}{\partial \,p_a\;\partial \,q^{\prime
\,\,j}\;}}\ \right| ^{\ 1/2}  \nonumber \\
&&\ \   \nonumber \\
&&\exp \left[ {\frac i\hbar }\ \left( J({\bf q}^{\prime \prime }\,t^{\prime
\prime }\,|\,{\bf p\,}t)+J({\bf p\,}t\,|{\bf q}^{\prime }\,t^{\prime
})\right) \right] \ ,  \label{(9)}
\end{eqnarray}
where $t$ is prescribed to be the mid time: $t\equiv t^{\prime }+(\epsilon
\,/2)=t^{\prime \prime }-(\epsilon \,/2)$. When $\epsilon =0$, one gets the
orthonormality relation between eigenstates of the operator $\hat {\bf q}$
(see Eqs.(\ref{(181)}) and (\ref{(182)})).

\bigskip

Since canonical coordinates and momenta were treated on an equal footing,
one realizes that the propagator in the $p_a$-representation which is
consistent with Eq.(\ref{(9)}) is

\begin{eqnarray}
{\cal K}({\bf p}^{\prime \prime }\ t^{\prime \prime }=t^{\prime }+\epsilon
\;|\;{\bf p}^{\prime }\ t^{\prime })=\int \frac{d^nq^j}{(2\pi \hbar )^n}{\ }
\ &&\left| \ \det -{\frac{\partial ^2J({\bf p}^{\prime \prime }\,t^{\prime
\prime }\,|\,{\bf q\,}t)}{\partial p_a^{\prime \prime }\;\partial \,q^{\,j}}}
\ \ \right| ^{\ 1/2}\ \ \left| \det \frac{\partial ^2J({\bf q\,}t\,|\,{\bf p}
^{\prime }\,t^{\prime })}{\partial \,q^j\;\partial \,p_a^{\prime }\ }\
\right| ^{1/2}\   \nonumber \\
&&  \nonumber \\
&&\ \exp \left[ {\frac i\hbar }\ \left( J({\bf p}^{\prime \prime }t^{\prime
\prime }|{\bf q}t)+J({\bf q}t|{\bf p}^{\prime }t^{\prime })\right) \right] \
.  \label{(90)}
\end{eqnarray}

\bigskip

\ 

It will become clear in Section V that the prescription of mid time in Eqs.( 
\ref{(9)}),(\ref{(90)}) together with the splitting of the Jacobian,
guarantee the hermiticity of the Hamiltonian operator and the unitarity of
the evolution.

\section{Classical systems}

To illustrate the use of Eq.(\ref{(9)}), let us consider a one-dimensional
classical system governed by the Hamiltonian

\begin{equation}
H\ =\ {\frac{p^2}{2 m}}\ +\ V(q).  \label{(40)}
\end{equation}

We will show how to manage the integration in Eq.({\ref{(9)}) in order to
get the infinitesimal propagator in the form of Eq.(\ref{(a0)}). }

{Since the metric in the Hamiltonian is a standard euclidean metric (}$
g^{ij}=\delta ^{ij}$){, the coordinate }$q$ is the normal coordinate. The
Jacobi principal function $J(q,p,\tau )$ can be guessed by writing

\begin{equation}
J(q,p,\tau )\ =\ p\ q\ -\ {\frac{p^2}{2m}}\ \tau \ +\ \sum_{l=1}^\infty
J_l(q,p)\ \tau ^l\,.  \label{(41)}
\end{equation}
Then one solves the Hamilton-Jacobi equation order by order in $\tau $, and
obtains

\begin{eqnarray}
&&J_1\ =\ - V(q)  \nonumber \\
&&J_2\ =\ {\frac{p\ V^{\prime}(q)}{2 m}}  \nonumber \\
&&J_3\ =\ -{\frac{p^2\ V^{\prime\prime}(q)}{6 m^2}}\ -\ {\frac{
V^{\prime}(q)^2}{6 m}}  \nonumber \\
&&J_4\ =\ {\frac{p^3\ V^{\prime\prime\prime}(q)}{24 m^3}}\ +\ {\frac{5}{24}} 
{\frac{p}{m^2}} V^{\prime}(q) V^{\prime\prime}(q)  \nonumber \\
&&J_5\ =\ -{\frac{1}{120}} {\frac{p^4}{m^4}} V^{\prime\prime\prime\prime}(q)
- {\frac{1}{15}} {\frac{p^2 V^{\prime\prime}(q)^2}{m^3}}\ -\ {\frac{3}{40}} {
\ \ \ \frac{p^2 V^{\prime}(q) V^{\prime\prime\prime}(q)}{m^3}}\ -\ {\frac{1}{
15 }} {\ \frac{V^{\prime}(q)^2 V^{\prime\prime}(q)}{m^2}}.  \label{(42)}
\end{eqnarray}
\bigskip

The recurrence formulae is

\begin{equation}
J_{l+1}\ =\ - {\frac{1}{2(l+1)m}}\ \sum_{k=0}^l {\frac{\partial J_k}{
\partial q}} {\frac{\partial J_{l-k}}{\partial q}},\ \ \ \ \ \ l\ge 1,
\label{(43)}
\end{equation}
where $J_0 \equiv p q$. Each $J_l$ is polynomical in $p$. Let us concentrate
on the higher degree contributions; their addition is

\begin{equation}
- \sum_{l=1}^\infty\ {\frac{1}{l!}} \left( - {\frac{p }{m}}\right)^{l-1}
V^{(l-1)}(q)\ \tau^l\ =\ {\frac{m}{p}}\ \int_q^{q - p \tau/m} V(q)\ dq.
\label{(44)}
\end{equation}

\bigskip

We are going to replace this result, and $\partial^2 J/\partial q\partial p
= 1 + {\cal O}(\epsilon^2)$, in the integrand of Eq.(\ref{(9)}). After the
substitution $p\rightarrow P\equiv p\epsilon/(2 m)$, one gets

\begin{eqnarray}
&&K(q^{\prime\prime}\ t^{\prime\prime}=t^{\prime}+\epsilon |q^{\prime}\
t^{\prime})\ =\ \left(\frac{m}{\pi\hbar\epsilon}\right)\ \exp\left[{\frac{i
m (\Delta q)^2}{2\hbar\epsilon}}\right]  \nonumber \\
&&\left\{\int dP \exp\left[- {\frac{i 2 m}{\hbar \epsilon}} \left(P - {\frac{
\Delta q}{2}}\right)^2 - {\frac{i\epsilon}{2 \hbar P}} \left(\int_{q^{\prime
\prime}-P}^{q^{\prime\prime}} V(q)\ dq + \int_{q^{\prime}}^{q^{\prime}+P}
V(q)\ dq\right) + ...\right]\ +\ {\cal O}(\epsilon^2)\right\}.  \label{(45)}
\end{eqnarray}

\medskip

The contribution of the potential to the phase will be expanded about $P =
\Delta q/2$:

\begin{eqnarray}
{\frac{1}{2 P}} &&\left(\int_{q^{\prime\prime}-P}^{q^{\prime\prime}} V(q)\
dq + \int_{q^{\prime}}^{q^{\prime}+P} V(q)\ dq\right)  \nonumber \\
&&=\ \bar V\ -\ {\frac{2}{\Delta q}}\ \bar{\Delta V}\ \left(P - {\frac{
\Delta q }{2}}\right)\ +\ {\frac{4}{(\Delta q)^2}}\ \bar{\Delta V}\ \left(P
- {\ \frac{\Delta q}{2}}\right)^2\ +\ ...\ ,  \label{(346)}
\end{eqnarray}

where

\begin{equation}
\bar V\ \equiv\ {\frac{1}{\Delta q}}\ \int_{q^{\prime}}^{q^{\prime\prime}}
V(q)\ dq,\ \ \ \ \ \ \ \ \ \ \ \ \ \ \ \bar{\Delta V}\ \equiv\ \bar V\ -\
V\left({\frac{q^{\prime}+ q^{\prime\prime}}{2}}\right).  \label{(347)}
\end{equation}

Those contributions that were not explicitly written in Eq.(\ref{(45)}) can
be controlled by means of the result\cite{gros}

\begin{equation}
\int dx\ \ x^{2\gamma}\ \exp\left[-{\frac{i 2 m}{\hbar \epsilon}}
x^2\right]\ \propto\ \left(\frac{\hbar \epsilon}{2 m}\right)^{\gamma + {\ 
\frac{1}{2}}}.  \label{(g)}
\end{equation}

Then the leading contribution to the integration (\ref{(9)}) is

\begin{equation}
K(q^{\prime\prime}\ t^{\prime\prime}=t^{\prime}+\epsilon |q^{\prime}\
t^{\prime})\ =\ \sqrt{\frac{m}{\ 2 i \pi \hbar \epsilon}}\ \exp\left[{\frac{
i }{\hbar}}\left({\frac{m (\Delta q)^2}{2 \epsilon}}\ -\ {\frac{\epsilon}{
\Delta q}}\int_{q^{\prime}}^{q^{\prime\prime}} V(q)\ dq\right)\right].
\label{(348)}
\end{equation}

The infinitesimal propagator (\ref{(348)}) has the form (\ref{(a0)}). In
fact the phase in Eq.(\ref{(348)}) solves the Hamilton-Jacobi equation in
each argument at order $\epsilon $ (for all values of $q^{\prime }$ and $
q^{\prime \prime }$)\footnote{
Although the phase in Eq.(\ref{(348)}) has the merit of being a complete
solution of the Hamilton-Jacobi equation at order $\epsilon $ --it is the
Hamilton principal function at that order--, the integration on $q$ in the
composition of infinitesimal propagators (Eq.(\ref{(27')})) will be not
sensitive to a replacement of $\bar V$ by $V((q^{\prime \prime }+q^{\prime
})/2)$, or $(V(q^{\prime })+V(q^{\prime \prime }))/2$, etc. as a consequence
of the result (\ref{(g)}).}. The Schr\"odinger equation is satisfied at the
lowest order in $\hbar \epsilon /m$.

\section{Operator ordering}

Each recipe to path integrate implies an operator ordering for the
Hamiltonian in the wave equation. Our interest in this Section is to find
the operator ordering associated with the infinitesimal propagator (\ref{(9)}
). Let us derive Eq.(\ref{(9)}) with respect to $\epsilon$, at $\epsilon = 0$
:

\begin{eqnarray}
i\ \hbar \ &&{\frac \partial {\partial \epsilon }}K(q^{\prime \prime }\
t^{\prime }+\epsilon \,|q^{\prime }\ t^{\prime })\Big\vert_{\epsilon
=0}=\int {\ \frac{dp}{2\pi \hbar }}\ \exp \left[ {\frac i\hbar }\ p\ \Delta
\sigma\right]  \nonumber \\
\nonumber \\
&&\ \left\{ {\frac{i\hbar }4}\ \left( -{\frac{\partial ^2H(q^{\prime \prime
},p)}{\partial q^{\prime \prime }\partial p}}+{\frac{\partial ^2H(q^{\prime
},p)}{\partial q^{\prime }\partial p}}\right) \ +\ {\frac 12}\left(
H(q^{\prime \prime },p)\ +\ H(q^{\prime },p)\right) \right\} \ ,
\label{(319)}
\end{eqnarray}
where the short-time approximation (\ref{(91')}) has been used.

\medskip\ 

Eq.(\ref{(319)}) is linear in $H$. If $H$ can be expanded in a power series,
then it will sufficient to handle the ordering for a Hamiltonian

\begin{equation}
H\ =\ q^m\ p^k.  \label{(45')}
\end{equation}
If $q$ is a normal coordinate, then

\begin{eqnarray}
i\ \hbar \ &&{\frac \partial {\partial \epsilon }}K(q^{\prime \prime }\
t^{\prime }+\epsilon \,|q^{\prime }\ t^{\prime })\Big\vert_{\epsilon
=0}=\int {\ \frac{dp}{2\pi \hbar }}\ \exp \left[ {\frac i\hbar }\ p\ \Delta
q\right]  \nonumber \\
\nonumber \\
&&\left\{ {\frac{i\hbar \ k\ m}4}\ p^{k-1}\ \left( -q^{\prime \prime }\
^{m-1}+q^{\prime }\ ^{m-1}\right) \ +\ {\frac{p^k}2}\left( q^{\prime \prime
}\ ^m+q^{\prime }\ ^m\right) \right\} .  \label{(49'')}
\end{eqnarray}
\medskip
Taking into account the Eq.(\ref{(92)}),

\begin{eqnarray}
\hat H \Psi(q^{\prime\prime},0) &&= \int {\frac{dq^{\prime}\ dp}{2\pi\hbar}}
\ \exp\left[{\frac{i}{\hbar}}\ p\ \Delta q\right]  \nonumber \\
\nonumber \\
&&\left\{{\frac{i\hbar\ k\ m}{4}}\ p^{k-1}\ \left(- q^{\prime\prime}\ ^{m-1}
+ q^{\prime}\ ^{m-1}\right)\ +\ {\frac{p^k}{2}} \left(q^{\prime\prime}\ ^m +
q^{\prime}\ ^m\right)\right\}\ \Psi(q^{\prime},0)  \nonumber \\
\nonumber \\
&&=\ -q^{\prime\prime}\ ^{m-1} {\frac{i\hbar\ k\ m}{4}}\ \left({\frac{\hbar}{
i}}{\frac{\partial}{\partial q^{\prime\prime}}}\right)^{k-1} \int {\frac{
dq^{\prime}\ dp}{2\pi\hbar}}\ \exp\left[{\frac{i}{\hbar}}\ p\ \Delta
q\right]\ \Psi(q^{\prime},0)  \nonumber \\
\nonumber \\
&&+\ {\frac{i\hbar\ k\ m}{4}}\ \left({\frac{\hbar}{i}}{\frac{\partial}{
\partial q^{\prime\prime}}}\right)^{k-1} \int {\frac{dq^{\prime}\ dp}{
2\pi\hbar}}\ q^{\prime}\ ^{m-1}\ \exp\left[{\frac{i}{\hbar}}\ p\ \Delta
q\right]\ \Psi(q^{\prime},0)  \nonumber \\
\nonumber \\
&&+\ {\frac{1}{2}}\ q^{\prime\prime}\ ^m \ \left({\frac{\hbar}{i}}{\frac{
\partial}{\partial q^{\prime\prime}}}\right)^k \int {\frac{dq^{\prime}\
dp}{2\pi\hbar}}\ \exp\left[{\frac{i}{\hbar}}\ p\ \Delta q\right]\
\Psi(q^{\prime},0)  \nonumber \\
\nonumber \\
&&+\ {\frac{1}{2}} \ \left({\frac{\hbar}{i}}{\frac{\partial}{\partial
q^{\prime\prime}}}\right)^k \int {\frac{dq^{\prime}\ dp}{2\pi\hbar}}\ \
q^{\prime}\ ^m\ \exp\left[{\frac{i}{\hbar}}\ p\ \Delta q\right]\
\Psi(q^{\prime},0)  \nonumber \\
\nonumber \\
&&=\ {\frac{i\hbar\ k\ m}{4}}\ \left[\hat p^{k-1},\hat q^{m-1}\right]\
\Psi(q^{\prime\prime},0)\ +\ {\frac{1}{2}}\left(\hat p^k\ \hat q^m \ +\ \hat 
q^m\ \hat p^k\right)\ \Psi(q^{\prime\prime},0).  \label{(59'')}
\end{eqnarray}

This means that the Hamiltonian operator is

\begin{equation}
\hat H\ =\ {\frac{1}{2}}\left(\hat p^k\ \hat q^m \ +\ \hat q^m\ \hat p
^k\right)\ +\ {\frac{i\hbar\ k\ m}{4}}\ \left[\hat p^{k-1},\hat q
^{m-1}\right].  \label{(60)}
\end{equation}

The operator $\hat H$ is Hermitian thanks to the mid time prescription in
Section III, which gave an equal weight to the terms depending on $
q^{\prime} $ and $q^{\prime\prime}$ in Eq.(\ref{(319)}).

\section{Conclusions}

We have proposed a scheme to path integrate in phase space, which is
applicable to Hamiltonian systems whose configuration space is a manifold
where normal coordinates (i.e., coordinates behaving like the components of
a vector in the tangent space at the origin) can be introduced. The
skeletonization is based on the invariant Jacobi principal functions
---those related with the variational principles of Mechanics for mixed
boundaries left fixed---, and the measure gives to the propagator the
character of a density of weight 1/2 in each argument. The so obtained
infinitesimal propagators (\ref{(9)}) and (\ref{(90)}) naturally satisfy the
Schr\"odinger equation, once the Hamiltonian operator is build in agreement
with the operator ordering (\ref{(60)}) induced by the path integral recipe.

The infinitesimal propagators (\ref{(9)}) and (\ref{(90)}) can be read in
terms of the modes

\begin{equation}
{\cal F}_{{\bf p}}(q^j,t)\ \equiv \ \left[ \det \left( {\frac 1{2\pi \hbar }}
\ {\frac{\partial ^2J({\bf q},{\bf p},t)}{\partial \,q^j\;\partial \,p_a}}
\right) \right] ^{1/2}\ \exp \left[ {\frac i\hbar }J({\bf q},{\bf p}
,t)\right] ,  \label{(53)}
\end{equation}
which are well behaved on all phase space at short times (the matrix $
\partial ^2J/\partial q^j\partial p_a$ is not singular because the relation
between initial and final momenta is biunivoque at short times). At $t=0$
these modes are eigenfunctions of the momenta $\hat p_a=-i\hbar \partial
/\partial \sigma ^a$ based at the origin $O$, because of the boundary
condition $J({\bf q},{\bf p},t=0)=p_a\sigma ^a$. Therefore ${\cal F}_{{\bf p}
}(q^j,t)$ is a short-time approximation for $<q^j\,|\hat U(t)|\,p_a>$, and $
\{{\cal F}_{{\bf p}}\}$ is a basis of short-time solutions of the
Schr\"odinger equation whatever the Hamiltonian system is. A change of the
origin $O$ implies a change of the basis $\{{\cal F}_{{\bf p}}\}$; of
course, all bases $\{{\cal F}_{{\bf p}}\}$ are equally good for expanding
the propagator.

For free systems it is $J({\bf q},{\bf p},t)=p_a\sigma ^a{\bf \ }-H(p_a)\,t$
, and the modes (\ref{(53)}) are certainly exact solutions of the
Schr\"odinger equation. They span the basis of eigenstates of the
(conserved) momenta $\hat p_a$. In this case the propagator (\ref{(9)}) is
exact (i.e., it is the finite propagator). In particular, the Newton-Wigner
propagator (\ref{(a7)}) for the relativistic particle --$
H(p)=(p^2+m^2)^{1/2} $-- can be obtained by integrating on the momenta in
Eq.(\ref{(9)})\cite{fe}.

\bigskip

In the case of the classical system of Section IV, the modes ${\cal F}_p$
satisfy the equation (use Eq.(\ref{j1}))

\begin{equation}
\left( i\hbar {\frac \partial {\partial t}}+{\frac{\hbar ^2}{2m}}{\frac{
\partial ^2}{\partial q^2}}-V(q)\right) \ {\cal F}_p\ =\ {\frac{\hbar ^2}{2m}
}\ \left( \frac{\partial ^2J}{\partial q\partial p}\right) ^{-1/2}\ {\frac{
\partial ^2}{\partial q^2}}\left[ \left( \frac{\partial ^2J}{\partial
q\partial p}\right) ^{1/2}\right] \ {\cal F}_p.  \label{(54)}
\end{equation}
This equation is typical for any phase $J$ being a solution of the
Hamilton-Jacobi equation, and is commonly used to remark the semiclassical
character ($\hbar \rightarrow 0$) of wavefunctions having the form (\ref
{(53)}). However, as was already stated, the Jacobi principal function 
$J({\bf q},{\bf p},t)$ confers an additional property to the modes (\ref
{(53)}) --which is the one exploited in this paper-- and allows a different
reading of Eq.(\ref{(54)}): the modes ${\cal F}_p$ are short-time solutions
of the Schr\"odinger equation, {\it for any value of $\hbar $}. In fact, the
results in Section IV show us that the rhs in Eq.(\ref{(54)}) is zero for
a quadratic potential\footnote{
If $V(q)=m \omega^2 q^2/2$, the Jacobi principal function is $J(q, p, t) =
- {p^2 \tan(\omega t)/(2 m \omega)} + {q p/\cos\omega t} - m \omega q^2
\tan(\omega t)/2$. Thus $J(q,p,t)$ reflects the equivalent roles of $q$ and 
$p$ in both the initial condition and the Hamiltonian.} (${\cal 
F}_p$ is an exact solution), $\hbar ^2t^4/(8m^3)(V^{\prime \prime \prime })^2
{\cal F}_p+{\cal O}(t^5)$ for a cubic potential, and $\hbar ^2\ t^2/(8\
m^2)V^{\prime \prime \prime \prime }{\cal F}_p\ +\ {\cal O}(t^3)$ in a more
general case.

\bigskip\ 

The substitution of the functional action by a skeletonized version in the
discrete-time approximation is one of the ways to give a meaning to the
functional integration (\ref{(1)}). A different approach to the same problem
is the operator symbol method \cite{ber}, where the path integral results
from the product of the {\it symbols} associated with the short-time
evolution operator. The product of symbols involves an integration
containing the information about the operator ordering, which amounts to the
prescription of the skeletonization in phase space. The different rules to
generate the ordering for the quantum operator ${\hat g}$ associated with a
function $g(q,p)$ in phase space, can be summarized as follows\cite{cohen}:

\begin{equation}
\hat g\ \ \ \longleftrightarrow \ \ \ \ f\left( -i{\frac \partial {\partial q
}},-i{\frac \partial {\partial p}}\right) \ \ \exp \left( -{\frac{i\hbar }2}{
\frac{\partial ^2}{\partial q\partial p}}\right) \ \ g(q,p),  \label{(77)}
\end{equation}
where $\hat g$ in Eq. (\ref{(77)}) is the normal form of the operator (the
power series expansion where the $q$ precede the $p$), and $f(u,v)$
contains the information about the ordering. Some well known rules of
ordering are\cite{grosche}:

\medskip\ 
\begin{tabular}{|c|c|c|} \hline
\text{{\bf Name}} & {$f(u,v)$} & {$g(q,p)=q^m\,p^k$}\\ \hline
\text{{\it Standard}} &   $\exp [\frac i2\hbar uv]$   & $\hat q^m\,\hat p^k$ \\ 
&  &    \\
\text{{\it Anti-Standard}}   & $\exp [-\frac i2\hbar uv]$  & $\hat p^k\,
\hat q^m$ \\ 
&  &    \\ 
\text{{\it Symmetric}} &   $\cos [\frac 12\hbar uv]$   & $\frac 12(\hat q
^m\,\hat p^k+\hat p^k\,\hat q^m)$ \\ 
&  &    \\ 
\text{{\it Weyl}}   & $1$   & $\;\;\frac 1{2^m}\sum\limits_{l=0}^m
{m \choose l}
\hat q^{m-l}\,\hat p^k\,\hat q^l\;\;$ \\ 
&  &    \\ 
\text{{\it Born-Jordan}} &   $\quad 2\,(\hbar uv)^{-1}\sin [\frac 12\hbar
uv]\quad$    & $\;\;\frac 1{k+1}\sum\limits_{l=0}^k\;\hat p^{k-l}\,\hat q^m\,
\hat p^l\;\;$ \\
\hline
\end{tabular}

\bigskip\ 

The Weyl ordering is equivalent to a skeletonization where the Hamiltonian
is evaluated in $(q^{\prime \prime }+q^{\prime })/2$. The symmetric ordering
corresponds to replace the Hamiltonian by $[H(q^{\prime \prime
},p)+H(q^{\prime },p)]/2$\cite{cohen2}. In this sense, the skeletonization
prescribed in this paper seems to be related with the symmetrization rule
(see Eq.(\ref{(91')})). Efectively, the ordering (\ref{(60)}) for the normal 
coordinates and their conjugated momenta begins with a symmetrized contribution 
coming from the mean Hamiltonian in Eq.(\ref{(319)}). However, our prescription 
includes a non-trivial measure (the Jacobians in Eq.(\ref{(76)})), which is 
needed in order that the wavefunction retains its condition of being a density 
of weight 1/2. This measure is responsible for the second term in the ordering 
(\ref{(60)}). Thus the function $f(u,v)$ associated with the ordering 
(\ref{(60)}) is

\begin{equation}
f(u,v)=\cos [\frac 12\hbar uv]+\frac 12\hbar uv\;\sin [\frac 12\hbar uv].
\label{78}
\end{equation}
Naturally $f$ fulfills the requirements 

\begin{equation}
\lim_{\hbar\rightarrow 0}f = 1\;\;\;\;\;\;\; \lim_{\hbar\rightarrow 
0}\dot f = 0   \label{79}
\end{equation}
that guarantee the classical limit \cite{cohen}.

\newpage

\noindent{\bf ACKNOWLEDGMENTS}

The author wishes to thank K.V.Kucha\v r for his comments on an early
version of this manuscript, and F.Gaioli, E.Garc\'\i a Alvarez and
M.Thibeault for helpful discussions. This work was supported by Universidad
de Buenos Aires (Proy. TX64) and Consejo Nacional de Investigaciones
Cient\'\i ficas y T\'ecnicas (Argentina).


\vskip1cm

\begin{references}
\bibitem{d}  P.A.M.Dirac, Physik. Zeits. Sowjetunion {\bf 3} (1933), 64.

\bibitem{f}  R.P.Feynman, Rev.Mod.Phys. {\bf 20} (1948), 367.

\bibitem{albe}  S.Albeverio, in {\it Proceedings of the N.Wiener Centenary
Congress (Michigan State University, 1994)}, ed. by V. Mandrekar et al, 
Proc. Appl. Math. {\bf 52} (AMS Providence RI, 1977).

\bibitem{dw}  B.S.DeWitt, Rev.Mod.Phys. {\bf 29} (1957), 377.

\bibitem{fh}  R.P.Feynman and A.R.Hibbs, {\it Quantum Mechanics and Path
Integrals}, McGraw-Hill, New York (1965)

\bibitem{sch}  L.S.Schulman, {\it Techniques and Applications of Path
Integration}, J.Wiley, N.Y. (1981).

\bibitem{mor}  C.Morette, Phys.Rev. {\bf 81}, 848 (1951).

\bibitem{vv}  J.H.Van Vleck, Proc. Natl. Acad. Sci. USA {\bf 14} (1928), 178.

\bibitem{fe}  R.Ferraro, Phys. Rev. D {\bf 45} (1992), 1198.

\bibitem{and}  A.Anderson, Phys. Rev. D {\bf 49} (1994), 4049.

\bibitem{k}  K.Kucha{\v r}, J.Math.Phys. {\bf 24} (1983), 2122.

\bibitem{p}  L.Parker, Phys. Rev. D {\bf 19} (1979), 438.

\bibitem{fi}  P.P.Fiziev, Theor.Math.Phys. {\bf 62} (1985), 123; also in 
{\it Lectures on Path Integration, Trieste 1991}, eds. H.Cerdeira et al,
W.Scientific, Singapore (1993), pp. 556-562.

\bibitem{ll}  L.D.Landau and E.M.Lifshitz, {\it Mechanics}, Pergamon Press,
Oxford (1959).

\bibitem{lanc}  C.Lanczos, {\it The Variational Principles of Mechanics},
Dover, New York (1986).

\bibitem{schutz}  B.F.Schutz, {\it Geometrical Methods of Mathematical
Physics}, Cambridge University Press, Cambridge (1980).

\bibitem{gros}  See, for instance, H.Kleinert, {\it Path integrals in
Quantum Mechanics, Statistics and polymer physics}, World Scientific,
Singapore (1995), or C.Grosche, {\it An introduction into the Feynman path
integral}, hep-th/9302097, pp. 14-15.

\bibitem{ber}  F.Berezin, Sov.Phys.Usp. {\bf 23} (1980), 763.

\bibitem{cohen}  L.Cohen, J.Math.Phys. {\bf 7} (1966), 781.

\bibitem{grosche}  C.Grosche, {\it ibidem}, p.8.

\bibitem{cohen2}  L.Cohen, J.Math.Phys. {\bf 11} (1970), 3296.
\end{references}
\end{document}